\documentclass[aps,pra,twocolumn,superscriptaddress]{revtex4-1}
\usepackage{amssymb,amsmath}
\usepackage{hyperref}
\usepackage{graphicx}

\DeclareMathOperator{\diag}{diag}
\DeclareMathAlphabet{\mathpzc}{OT1}{pzc}{m}{it}

\begin{document}

\title{Raman scattering with strongly coupled vibron-polaritons}

\author{Artem Strashko}
\affiliation{SUPA, School of Physics and Astronomy, University of St Andrews, St Andrews, KY16 9SS, United Kingdom}
\author{Jonathan Keeling}
\affiliation{SUPA, School of Physics and Astronomy, University of St Andrews, St Andrews, KY16 9SS, United Kingdom}
\date{\today}

\begin{abstract}
  Strong coupling between cavity photons and molecular vibrations can
  lead to the formation of vibron-polaritons. In a recent experiment
  with PVAc molecules in a metal-metal microcavity [A.~Shalabney {\it
    et al.}, Ang.~Chem.~Int.~Ed.  {\bf 54} 7971 (2015)], such a
  coupling was observed to enhance the Raman scattering probability by
  several orders of magnitude.  Inspired by this, we theoretically
  analyze the effect of strong photon-vibron coupling on the Raman
  scattering amplitude of organic molecules.  This problem has
  recently been addressed in [J.~del Pino, J.~Feist and
  F.~J.~Garcia-Vidal; J.~Phys.~Chem.~C {\bf 119} 29132 (2015)] using
  exact numerics for a small number of molecules. In this paper we
  derive compact analytic results for any number of molecules, also
  including the ultra-strong coupling regime. Our calculations predict
  a division of the Raman signal into upper and lower polariton modes,
  with some enhancement to the lower polariton Raman amplitude due to
  the mode softening under strong coupling.
\end{abstract}

\maketitle

\section{Introduction}

Light can be used to probe condensed matter systems, but also, as is
increasingly being explored, light can be used to change the material
properties of systems.  Examples of the latter range from topological Floquet
insulators\cite{oka09,inoue10,kitagawa10,lindner11}, where electronic band
structure is modified by a drive field, to light induced
superconductivity~\cite{fausti11,mankowsky14,denny15,mitrano16}.  These
examples all rely on strong driving, however recently there have been
experimental~\cite{schwartz11,hutchison12,orgiu15} and
theoretical~\cite{cwik14,galego15,michetti15,spano15,feist15,cwik16} works exploring how similar
effects can arise without driving for organic materials strongly coupled to optical
microcavities.  In some cases, light can be used both to probe the system, as
well as to change its properties.  This applies particularly when there are
multiple optically active transitions, such as infra-red active vibrational
modes in addition to optical frequency electronic
transitions~\cite{shalabney15:raman}.  This paper studies such a problem in
detail.

Organic materials are excellent systems for the exploration of strong
matter-light coupling, due to their large electronic oscillator strengths and
high binding energies. Most work has focused on strong coupling of light to
electronic transitions~\cite{lidzey98, lidzey99, lidzey00, tischler05,
  kena-cohen08} and the resultant formation of two hybrid matter-light
excitations, known as exciton-polaritons.  The strength of the matter-light
coupling can be characterized by the energy splitting between these modes.
Strong coupling occurs when this splitting exceeds the linewidth.
Ultra-strong coupling occurs when this splitting approaches the bare exciton
and photon energies~\cite{ciuti05}.  For organic exciton-polaritons, Rabi
splittings of 32\%~\cite{schwartz11},
52\%~\cite{liu15}, and up to 60\%~\cite{gambino14} of the bare exciton energy
have been demonstrated.  In addition to the interest arising from ultra-strong
coupling, organic materials are also interesting because of the relatively
strong coupling between electronic state and internal mechanical degrees of
freedom of organic molecules (rotations and vibrations), leading to the complex
interplay between matter-light coupling and internal structure discussed
above~\cite{schwartz11,hutchison12,cwik14,galego15,michetti15,orgiu15,spano15,feist15,cwik16}.
Of specific relevance to this paper, it was shown in several recent
experiments~\cite{shalabney15,george15,muallem15,simpkins15,long15}
that it is also possible to achieve strong coupling between infra-red
microcavities and vibrational modes of molecules, leading to
``vibron-polaritons''.

Organic materials where both electronic and vibronic transitions couple to
light, as well as coupling to each other, present rich possibilities for
manipulating properties of matter with light or matter-light coupling.  An
example of this was work by \citet{shalabney15:raman} where it was shown
experimentally that in an infra-red cavity, forming vibron-polaritons, there
were dramatic consequences for the Raman scattering (RS) of optical frequency
light.  The Raman transition probability to a final vibrationally excited
state splits between the vibron-polariton modes (referred to below as lower
polariton (LP) and upper polariton (UP)).  The most intriguing result of
\cite{shalabney15:raman} is however that the total Raman cross-section was
enhanced by three orders of magnitude when the infra-red cavity was resonant
with the vibrational modes.  Consequently, a new mechanism for RS enhancement
was proposed, which is essentially distinct from other methods of RS
enhancement such as stimulated RS~\cite{hummer15}, surface enhanced
RS~\cite{nie97,campion98}, or the recently proposed enhancement by parametric
plasmon-vibron coupling~\cite{roelli16}. 

Motivated by these experiments, the aim of this paper is to analyze the effect
of strong photon-vibron coupling on the RS probability.  In modeling organic
systems, a variety of approaches are possible~\cite{barford13,michetti15},
depending on the scale of the problem to be tackled.  In this paper, we are
focused on understanding the behavior of the $N$-molecule system for
arbitrary $N$, in order to explore what if any collective enhancement of Raman
scattering arises.  As such, we consider a simplified model of each molecule,
describing only one (harmonic) vibrational mode coupled to the electronic
transition.  Without further approximation, it is in fact possible to derive
exact formulae for Raman transition amplitudes.  The results we find could
also be generalized to multiple vibrational modes (while retaining a closed
form analytic expression), or to non-harmonic modes (but then losing the
closed form).  Given our aim of exploring the nature of collective
enhancement, such modifications of our model are not important.

We should note that theoretical calculations of Raman scattering with strongly
coupled vibron-polaritons has recently been addressed by~\citet{pino15a}, 
who discussed the general behaviour for $N$ molecules when treated as
three-level systems, and performed exact numerics for a small number of 
molecules using the same model we use below.  Their
results suggested there is no collective enhancement of Raman scattering. We
confirm and extend these results by presenting analytic results for an
arbitrary number of molecules, hence confirming the absence of a collective RS
enhancement effect.  We do however find that the total Raman amplitude can in
principle by significantly enhanced at ultra-strong coupling, by softening of
the lower polariton mode, however this requires coupling strengths in excess
of those seen in Ref.~\cite{shalabney15:raman}.

The remainder of this paper is organized as follows.  We divide our discussion
into calculations within the rotating wave approximation
(Section~\ref{sec:within-rotating-wave}) and beyond the rotating wave
approximation (Section~\ref{sec:ultra-strong-coupl}).
Section~\ref{sec:model-matr-elem} defines our notation, by presenting the
model we consider, and the matrix elements we must calculate.
Section~\ref{sec:calc-matr-elem} derives the explicit form of Raman transition
matrix elements as a sum over intermediate states.  Crucially,
section~\ref{overlap} then shows how these sums can be performed analytically,
resulting in a relatively compact expression.  Using coefficients and energies
derived in Sec.~\ref{sec:calc-eigenst}, section~\ref{sec:numerical-results}
presents numerical results, and analytic forms for the far detuned limit.
Beyond the rotating wave approximation, Section~\ref{sec:calc-matr-elem-1}
presents an alternate approach to calculating Raman transition matrix
elements, and Section~\ref{sec:numer-results-large} presents corresponding
numerical results.  Finally, in section~\ref{sec:multiple-excitations} we
extend the rotating-wave approximation formulae to consider final states with
multiple vibron-polaritons, and discuss the relative scaling with system size
of the different excitation number sectors.  Appendices provide further
details of some of the mathematical steps.

\section{Within the rotating wave approximation}
\label{sec:within-rotating-wave}

\subsection{Modeling Raman probabilities}
\label{sec:model-matr-elem}

\begin{figure}[!h]
\center{\includegraphics[width=3.2in]{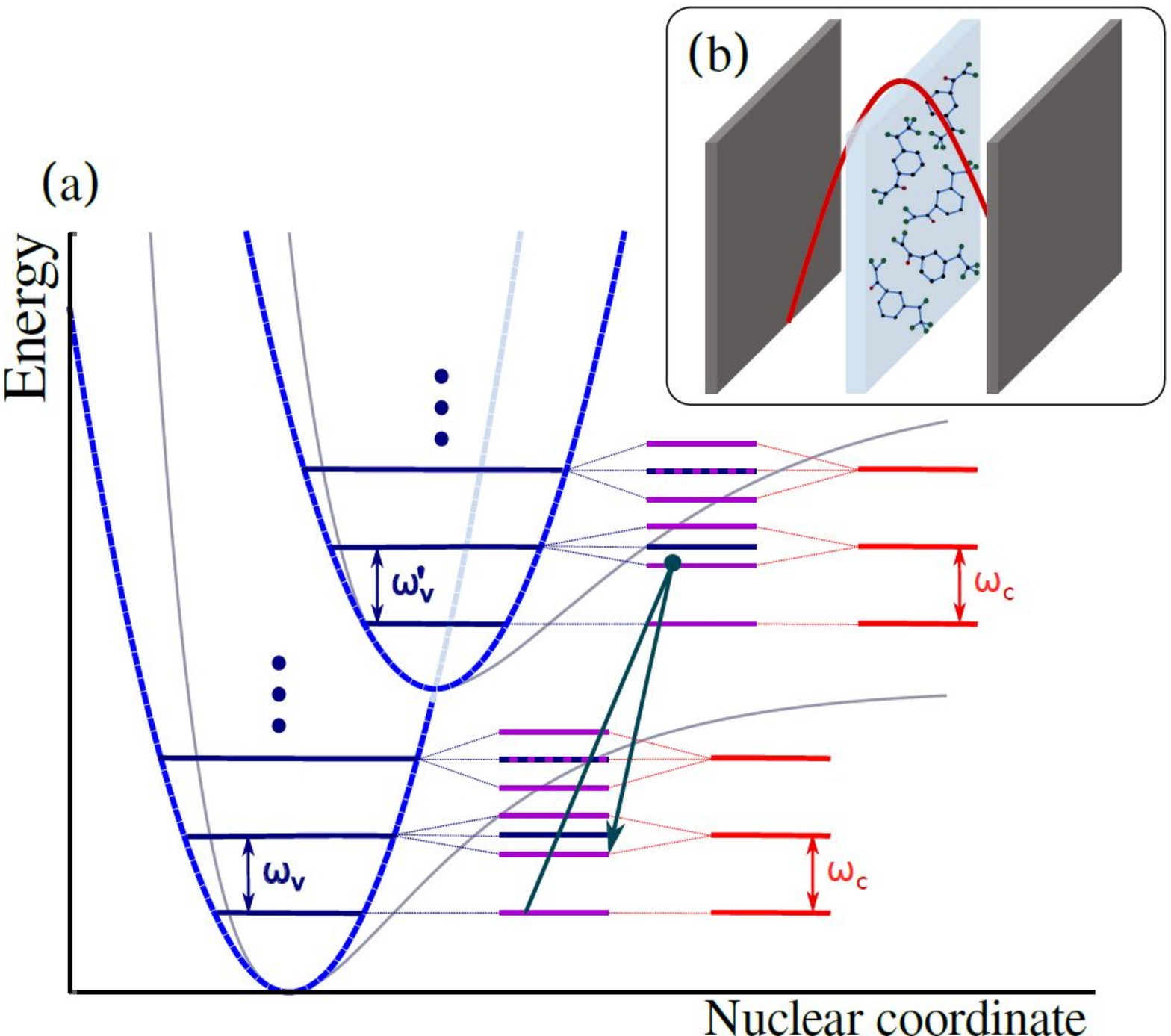}}
\caption{(a) Schematic illustration of the vibronic energy levels in
  the two electronic manifold (left) and their hybridization with
  photon number states (right) to form a ladder of polariton states
  (middle).  (b) Cartoon of molecules placed at the antinode of an
  optical cavity.}
\label{cartoon}
\end{figure}

We consider a single mode cavity, containing $N$ molecules.  We
represent each molecule by two degrees of freedom: two electronic
states (corresponding to HOMO and LUMO) levels, and a single vibrational
mode.
In this respect the model is similar to the ``Tavis-Cummings-Holstein''
model used recently~\cite{cwik14,cwik16,herrera15} to model vibrational dressing
of polaritons.  However, here we consider the case where it is the
molecular vibrations, rather than the electronic transition, which couples
to the cavity mode.
This model is shown schematically in Fig.~\ref{cartoon}.

The main simplifying assumption in such a model (the same model as
used in Ref.~\cite{pino15a}) is the replacement of full intramolecular
potential by a single harmonic degree of freedom. This is valid in the
limit where only a single collective mode dominates the physics,
either due to coupling most strongly to the electronic transitions, or
due to the resonant cavity coupling predominantly to one mode.
From the Raman spectrum seen without strong coupling, this is 
clearly the relevant regime in Ref.\cite{shalabney15:raman}.  In
this paper we will consider this problem both with and without the
rotating wave approximation (RWA). Within the RWA, the Hamiltonian
takes the following form:
\begin{multline}
\label{hamilt1}
H=\omega_c \hat a^{\dagger} \hat a + \sum_n \Bigl(  \omega_e \sigma_n^\uparrow
 + \omega_v [\hat b_n^{\dagger} \hat b^{\mathstrut}_n + 
\sqrt{S}(\hat b_n^{\dagger} + \hat b^{\mathstrut}_n)\sigma_n^\uparrow] + \\ +
G(\hat b_n^{\dagger}\hat a + \hat b^{\mathstrut}_n\hat a^{\dagger}) \Bigr).
\end{multline}
Here $\hat a$ is the annihilation operator for the cavity photon modes
with frequency $\omega_c$. The Pauli operators $\sigma_n$ describe
transitions of the electronic state of molecule $n$, with energy
splitting $\omega_e$, and we have used the shorthand
$\sigma^\uparrow_n = (1+\sigma^z_n)/2$ for the projector onto the excited state.
Finally $\hat b_n$ is the annihilation operator for
the vibrational mode of molecule $n$, with frequency $\omega_v$.  The
coupling between electronic and vibrational states is parameterized by
the Huang--Rhys parameter $S$, which describes the relative
displacement of the vibrational mode between the electronic ground
and excited states.  The coupling between cavity photons and
vibrational modes is denoted $G$.  

Using the above Hamiltonian, we are going to calculate the probability
of Raman scattering to a polaritonic mode (in the presence of a
cavity) and compare it with the Raman scattering probability to the
bare vibrational mode without a cavity.  In order to study Raman
scattering, we consider a weak driving field $E_{applied}(t) \sum_n \sigma_n^x
$, which we treat perturbatively.  In second order
perturbation theory, and using the resonant approximation the
probability of scattering can be written as~\cite{berestetskii71}:
\begin{equation}
\label{mostgeneralprob}
P_{0\rightarrow f_k} = \left| \sum_P \frac{\langle 0|\hat V_1|P \rangle\langle P|\hat V_2|f_k \rangle}{E_P-E_0-\omega} \right|^2,
\end{equation}
where $\omega$ is the frequency of the applied probe field, $E_0$ is
initial state energy and $E_P$ the intermediate state energy.  The
states $|0\rangle, |P\rangle, |f_k\rangle$ denote initial, intermediate
and final states --- we have allowed a label $k$ to distinguish
different final states (e.g.\ upper vs lower polariton excitations).
The operators $\hat V_1, \hat V_2$ can be written explicitly in terms
of coupling between the total dipole operator $\sum_m \sigma_m^x$
of the molecules
and the incident and emitted light. As our aim is ultimately to
compare the probabilities for polaritonic and "ordinary" Raman
scattering, we can however ignore all constant prefactors.  Ignoring
also dependence on the polarization of the light we may write the
transition probability as:
\begin{equation}
P_{0 \rightarrow f_k} = \gamma \left| M_k \right|^2,
\quad
\label{mk}
M_k = \sum_{m,P} \frac{\langle 0| \sigma_m^-|P\rangle\langle P | \sigma_m^+|f_k\rangle}{E_P - E_{0} - \omega}
\end{equation}
where $\gamma$ describes the (constant) electronic matrix elements and
density of final photon states, $m$ labels the specific molecules that
is excited, and $P$ labels the intermediate states.  NB, the sum over
molecules appears within the modulus squared, so that interference between
separate molecules' Raman scattering processes are allowed.
Note also that in Eq.~(\ref{mk}) there are no cross terms between
different molecules, as these vanish due to the assumed initial electronic 
ground state.

When considering the experimentally measured Raman spectrum, this can 
written as corresponding to:
\begin{equation}
\label{sumdegen}
P(\nu)\propto \sum_k\delta(\nu-E_k)|M_k|^2
\end{equation}
where $E_k$ is the energy of the final state mode, and $\nu$ is the
measured stokes shift.  This can be important when multiple degenerate
modes exist, such that the labeling of final states is arbitrary.  In
such a case, the measurable quantity is the sum of the probabilities
of transitions to the manifold of degenerate final states.

\subsection{Calculating matrix elements}
\label{sec:calc-matr-elem}

Calculating the amplitude $M_k$ in Eq.~(\ref{mk})  requires us to
find the initial, intermediate and final eigenstates of
Eq.~(\ref{hamilt1}), and evaluate the matrix elements of
$\sigma_n^\pm$ between these states.  Since Eq.~(\ref{hamilt1}) is
clearly diagonal in electronic state, there are two cases we should
consider, the electronic ground state, which we denote
$H_{\text{eff},\Downarrow}$ and the state where the $m^{\text{th}}$ molecule
is electronically excited, $H_{\text{eff},m}$
\begin{align}
\label{eq:heff0}
H_{\text{eff},\Downarrow}&=
\omega_c \hat a^{\dagger}\hat a + 
\sum_n \left[
\omega_v \hat b_n^{\dagger}\hat b^{\mathstrut}_n 
+ G(\hat b_n^{\dagger}\hat a + \hat b^{\mathstrut}_n\hat a^{\dagger}) \right]
\\
\label{eq:heffm}
H_{\text{eff},m}&=
H_{\text{eff},\Downarrow}
+ \omega_v\sqrt{S}(\hat b_m^{\mathstrut} + \hat b^{\dagger}_m).
\end{align}

For the electronic ground state, $H_{\text{eff},\Downarrow}$ can be
diagonalized by introducing $ \hat \xi_i= \upsilon_i \hat a + \sum_{n}
U_{n,i} \hat b_n $, which obey the required commutation relations
$[\hat \xi_i,\hat \xi_j^{\dagger}]=\delta_{i,j}$.  In this
diagonalized basis we may write $H_{\text{eff},\Downarrow}=\sum_i
\omega_i \hat \xi_i^\dagger \hat \xi_i^{}$ where $\omega_i$ denotes
the frequencies of the normal modes.  These give us $N+1$ eigenmodes:
2 polaritonic modes and $N-1$ degenerate dark modes (which have no
photonic part, $\upsilon_i \equiv 0$).  From the permutation symmetry
of the Hamiltonian, it is clear that for the polaritonic modes
$U_{n,i \in LP,UP}$ should be independent of $n$, and so orthogonality requires
that the dark modes satisfy $\sum_n U_{n,i \in \text{Dark}} \equiv 0$.

For the excited state $H_{\text{eff},m}$, diagonalization requires an
additional linear displacement to remove the linear terms.
Since the quadratic terms in Eq.~(\ref{eq:heff0},\ref{eq:heffm}) are
identical, the unitary transformation required is the same for both
Hamiltonians.  This means one may write $\hat\eta_i = \hat\xi_i +
\alpha_{m,i}$, one may use the identity
\begin{align*}
H_{\text{eff},m}
&=
\sum_i \left[\omega_i \hat
\xi_i^\dagger \hat \xi_i^{} 
+ \omega_i \left( \alpha_{m,i}^\ast \hat \xi_i^{} + 
\alpha_{m,i}^{} \hat \xi_i^{\dagger} \right)\right]
\nonumber\\
&=\sum_i \omega_i \hat \eta_i^\dagger \hat \eta_i^{} - \omega_i|\alpha_{m,i}|^2.
\end{align*}  
to diagonalize the problem. Comparison to Eq.~(\ref{eq:heffm}) shows that
this requires $\omega_i \alpha_{m,i} = U_{m,i} \omega_v \sqrt{S} $.  
Since the explicit form of the $\omega_i, \alpha_{m,i}$ is not required
to deriving the transition probability, we will defer its calculation
to section~\ref{sec:calc-eigenst}.  It is however useful to note that
from the above, we know that dark states, being purely vibrational
will have $\omega_i=\omega_v$ and obey $\sum_{m} \alpha_{m,i}=0$.

Using the linear relation between $\hat \eta_i$ and $\hat \xi_i$
given above, one may relate the ground state in the
electronic ground state manifold $
|0_\Downarrow\rangle=|\Downarrow;0_{LP},0_{UP},0_1,0_2,\ldots,0_{N-1}\rangle $
to that in the manifold where the $m$th molecule is excited $
|0_{m}\rangle=|\uparrow_m;0_{LP},0_{UP},0_1,0_2,\ldots,0_{N-1}\rangle $.  These
states are related by:
\begin{equation}
  \label{eq:offset}
|0_{m}\rangle = e^{-\sum_i \left( \alpha^{\mathstrut}_{m,i}\hat \xi_i^{\dagger} - \alpha_{m,i}^\ast\hat \xi^{\mathstrut}_i\right) } |0_\Downarrow\rangle.
\end{equation}

The matrix elements appearing in Eq.~(\ref{mk}) can then be written
out using this relation.  Let us denote the required overlaps as
$\mathpzc{M}^{(m)}_{0,P} \equiv \langle 0 | \sigma^-_m|P\rangle$ and
$\mathpzc{M}^{(m)}_{f_k,P} \equiv \langle f_k | \sigma^-_m|P\rangle$.
If we label the intermediate states $P$ by the set of occupations $\{p_i\}$ of
each normal mode, this expression becomes:
\begin{equation}
  \label{eq:Mkvsp}
  M_k =
  \sum_{m,\{p_i\}}
  \frac{\mathpzc{M}_{f_k,\{p_i\}}^{(m)\ast} \mathpzc{M}_{0,\{p_i\}}^{(m)}}{%
    \Delta + \sum_i p_i \omega_i}
\end{equation}
where $\Delta=\omega_e-\omega$ is the detuning of the
probe laser below the electronic transition.  Using the displacement
relation in Eq.~(\ref{eq:offset}), we may see what the overlap between
ground state and intermediate state is given by:
\begin{align}
\mathpzc{M}_{0,\{p_i\}}^{(m)}
&\equiv
\left< 0_{\Downarrow} \right|
\prod_i \frac{\hat \eta_i^{\dagger\; p_i}}{\sqrt{p_i!}}
\left| 0_{m} \right>
\nonumber\\
&= 
\left< 0_{\Downarrow} \right| \prod_i 
\frac{(\hat \xi_i^{\dagger} + \alpha^\ast_{m,i})^{p_i}}{\sqrt{p_i!}}
e^{-\alpha_{m,i}\hat\xi_i^\dagger - |\alpha_{m,i}|^2/2 } \left| 0_\Downarrow \right> 
\nonumber\\
&= 
\prod_i
\frac{\alpha_{m,i}^{\ast \; p_i}e^{- |\alpha_{m,i}|^2/2}}{\sqrt{p_i!}}.
\end{align}

The other matrix element describes the transition from the
intermediate state to a given final state.  If we consider the final state
with a single excitation of mode $k$, this can be written as:
\begin{equation}
  \label{oneExME}
\mathpzc{M}_{f_k,\{p_i\}}^{(m)}
=
\mathpzc{M}_{0,\{p_i\}}^{(m)}
\frac{p_k-|\alpha_{m,k}|^2}{\alpha^\ast_{m,k}}.
\end{equation}
As discussed in Section~\ref{sec:multiple-excitations} and
Appendix~\ref{sec:multiple-final-state}, this is a special case of the more
general formula for a final state with arbitrary occupations of multiple modes
in the final state.

Putting the above results together, we find the following expression 
for the matrix elements for single final-state excitations.
\begin{equation}
\label{sumcalc}
M_k =
\sum_{m} \frac{1}{\alpha_{m,k}}\sum_{\{p_i\}} 
\prod_{i} \left( 
  e^{-|\alpha_{n,i}|^2}\frac{|\alpha_{n,i}|^{2p_i}}{p_i!} 
\right)
\frac{p_k - |\alpha_{m,k}|^2}{\Delta + \sum_j p_j\varepsilon_j},
\end{equation}
In the limit of large $\Delta$, the denominator can be Taylor
expanded, and at leading order the summations can be
evaluated.  In the next section, we show that this can also be
rewritten in a form that makes its evaluation straightforward for all
parameter values.

\subsection{Compact form of matrix elements}
\label{overlap}

In evaluating the sum over $p_i$ in Eq.~(\ref{sumcalc}), the complication is
the appearance of $\sum_j p_j \epsilon_j$ in the denominator.  This can be
addressed by rewriting the denominator as the integral of an exponential,
which then allows all summations of $p_i$ to be evaluated analytically, as
follows:
\begin{multline*}
  M_k = 
  \int\limits_0^{\infty}dz e^{-z \Delta} \sum_{m} \frac{1}{\alpha_{m,k}}
  \sum_{\{p_i\}} \left( p_k - |\alpha_{m,k}|^2 \right)  
  \\
  \times
  \left( \prod_i \frac{\left( |\alpha_{m,i}|^2 e^{-z \omega_i} \right)^{p_i} e^{-|\alpha_{m,i}|^2}}{p_i!}
  \right)
  \\ =
  \int\limits_0^{\infty} e^{-z \Delta} \sum_{m} \frac{\left( e^{-z \omega_k} - 1 \right)}{\alpha_{m,k}}
  |\alpha_{m,k}|^2 \prod_i e^{|\alpha_{m,i}|^2 \left( e^{-z \omega_i} - 1 \right) },
\end{multline*}
thus we can  write the final expression in the compact form:
\begin{multline}
  \label{MkRWA}
  M_k = \sum_{m} 
  \int\limits_0^{\infty}dz e^{-z \Delta}  \alpha^{\ast}_{m,k}
  \left( e^{-z \omega_k} - 1 \right)  \\ 
  \times
  \exp \left[ -\sum_{i} |\alpha_{m,i}|^2 
    \left(1- e^{-z \omega_i} \right) \right].
\end{multline}
This is one of the central results of this manuscript; we next discuss
the analysis of this result, and then consider the generalization 
beyond the rotating wave approximation.

It can be immediately seen from Eq.~(\ref{sumdegen}) that there is no
transition to the dark modes, as orthogonality to bright states
implies that $\sum_m \alpha_{m,k}=0$; we discuss this further below.
For the remaining bright states, $\alpha_{m,k}$ is independent of $m$,
and so the sum over $m$ appearing in Eq.~(\ref{MkRWA}) can be replaced
by a factor $N$.  In the next section, we discuss further details of
the behavior of Eq.~(\ref{MkRWA}), which rely on the form of
$\omega_i, \alpha_{m,i}$.

\subsection{Calculating eigenstates}
\label{sec:calc-eigenst}

As noted above, $H_{\text{eff},\Downarrow}$ can be diagonalized by introducing
Bosonic operators $ \hat \xi_i= \upsilon_i \hat a + \sum_{n} U_{n,i} \hat
b_n$. This section discusses the coefficients $\upsilon_i, U_{n,i}$ and
frequencies $\omega_i$, which result.

The eigenstates divide into two classes; two polaritonic modes (involving
photons), and $N-1$ dark modes for which $\upsilon_i=0$.
For the polaritonic modes one has:
\begin{equation}
\omega_i\equiv\omega_{1,2}=\frac{\omega_c+\omega_v}{2}\pm \sqrt{\Bigl(\frac{\omega_c-\omega_v}{2}\Bigr)^2 + NG^2}.
\end{equation}
Enforcing Bosonic commutation relations on $\hat \xi_i$ determines
their normalization, so that for the two bright modes one may write:
$ U_{n,1} = \sin(\theta)/\sqrt{N}, \upsilon_1 = -\cos(\theta)$
and
$ U_{n,2} = \cos(\theta)/\sqrt{N}, \upsilon_2 = -\sin(\theta)$
where 
\begin{equation}
  \tan(2 \theta) = \frac{G \sqrt{N}}{(\omega_c-\omega_v)/2}.
\end{equation}
Note that for these modes, the symmetry of the matter-light coupling
requires that $U_{n,i}$ is independent of the molecule label $n$.

For the remaining $N-1$ dark modes ($\upsilon_i \equiv 0$) these are
purely vibronic and so $\omega_{i}=\omega_v$. Orthogonality to the
bright polaritons demands that $\sum_{n}U_{n,i} = 0$, and
normalization imposes the condition $\sum_{n}U_{n,i}U^\ast_{n,j}=
\delta_{i,j} $.

It is clear that the above equations do not uniquely define the dark-state
values of $U_{n,i}$; any $N-1$ orthonormal modes that are orthogonal to the
symmetric mode will suffice.  As such, the coefficients
\begin{equation}
\label{alphaRWA}
\alpha_{m,i} =
U_{m,i} \sqrt{S} \frac{\omega_v}{\omega_i}
\end{equation}
appearing in the observable Raman amplitude in Eq.~(\ref{MkRWA}) are not
uniquely determined.  However, as we discuss next, one can check that the
overall result of Eq.~(\ref{MkRWA}) is invariant under this freedom.

For all modes, the exponent involves the sum over all modes $\sum_{i}
|\alpha_{m,i}|^2 \left(1- e^{-z \omega_i} \right)$. Using
Eq.~(\ref{alphaRWA}), the contribution of dark modes to this sum can
be seen to be given by $ \sum_{i\in \text{Dark}} |U_{m,i}|^2
=(N-1)/N$, requiring only the orthonormality and completeness of the
coefficients $U_{m,i}$.  Since the bright modes have coefficients
$\alpha_{m,i}$ that are independent of the molecule label $m$, it is
clear that the exponent in Eq.~(\ref{MkRWA}) does not depend on the
molecule label $m$.  This confirms that the scattering rate into dark
modes vanishes because of the condition $\sum_{m} \alpha_{m,i \in
  \text{Dark}} = 0$, while for the bright modes, the sum over molecules
$m$ can be replaced by a factor $N$.

It is worth noting two explicit choices for $U_{m,i}$ that lead
to particularly simple demonstrations of the above results:

\paragraph{Symmetric dark-state basis.}
\label{sec:symmetric-basis}

The most obvious choice is to write 
\begin{equation}
  \label{UmatStd}
  U_{m,j}=\frac{\exp(i 2\pi m j/N)}{\sqrt{N}} 
\end{equation}
where $j=1 \ldots N-1$ for the dark modes.  This clearly satisfies
the above expressions as $|U_{m,j \in \text{Dark}}|^2 = 1$.  This choice
has the apparent advantage of treating all molecules
equivalently.  

\paragraph{Alternate dark-state basis.}
\label{sec:alternate-basis}

An alternate choice is to treat the molecule
$m$ that is electronically excited differently to the others.
This then leads to the choice:
\begin{equation}
  \label{UmatSep}
  \begin{split}
  U_{n, j_0} &= 
  \frac{1}{\sqrt{N(N-1)}}
  \begin{cases}
    N-1 & n = m \\
    -1  & n \neq m
  \end{cases}, \\
  U_{n,j \neq j_0} &=
  \frac{1}{\sqrt{N-1}} 
  \begin{cases}
    0 & n=m \\
    \exp\left(\frac{i 2\pi j \tilde{n}}{N-1}\right) & n \neq m
  \end{cases}
  \end{split}.
\end{equation} 
The quantity $\tilde{n}$ appearing in the last expression is a
sequential integer indexing the $N-1$ molecules excluding molecule $m$.
Note that there are only $N-2$ modes $j \neq j_0$ in the second expression
as $j$ and $j+N-1$ are equivalent, and $j=0$ is not orthogonal to
the mode $j_0$.

The advantage of this choice of basis is that $U_{m,j \neq j_0}=0$
means that these terms immediately drop out Eq.~(\ref{MkRWA}).  i.e.,
only three modes, two bright and one dark, contribute to the exponent.
For these three modes , one can write:
\begin{equation}
  \label{eq:ThreeModeAlpha}
  \alpha_{m,i}
  =
  \sqrt{\frac{S}{N}}
  \begin{pmatrix}
    \cos \theta  \frac{\omega_v}{\omega_{LP}}, 
    \sin \theta \frac{\omega_v}{\omega_{UP}}, 
    \sqrt{N-1}
  \end{pmatrix},
\end{equation}
and on resonance, one can further simplify $\cos \theta=\sin\theta =
1/\sqrt{2}$ and $\omega_{LP,UP} = \omega_v \mp G \sqrt{N}$.

\subsection{Numerical results and  large $\Delta$ approximation}
\label{sec:numerical-results}

\begin{figure}[!h]
\center{\includegraphics[width=3.2in]{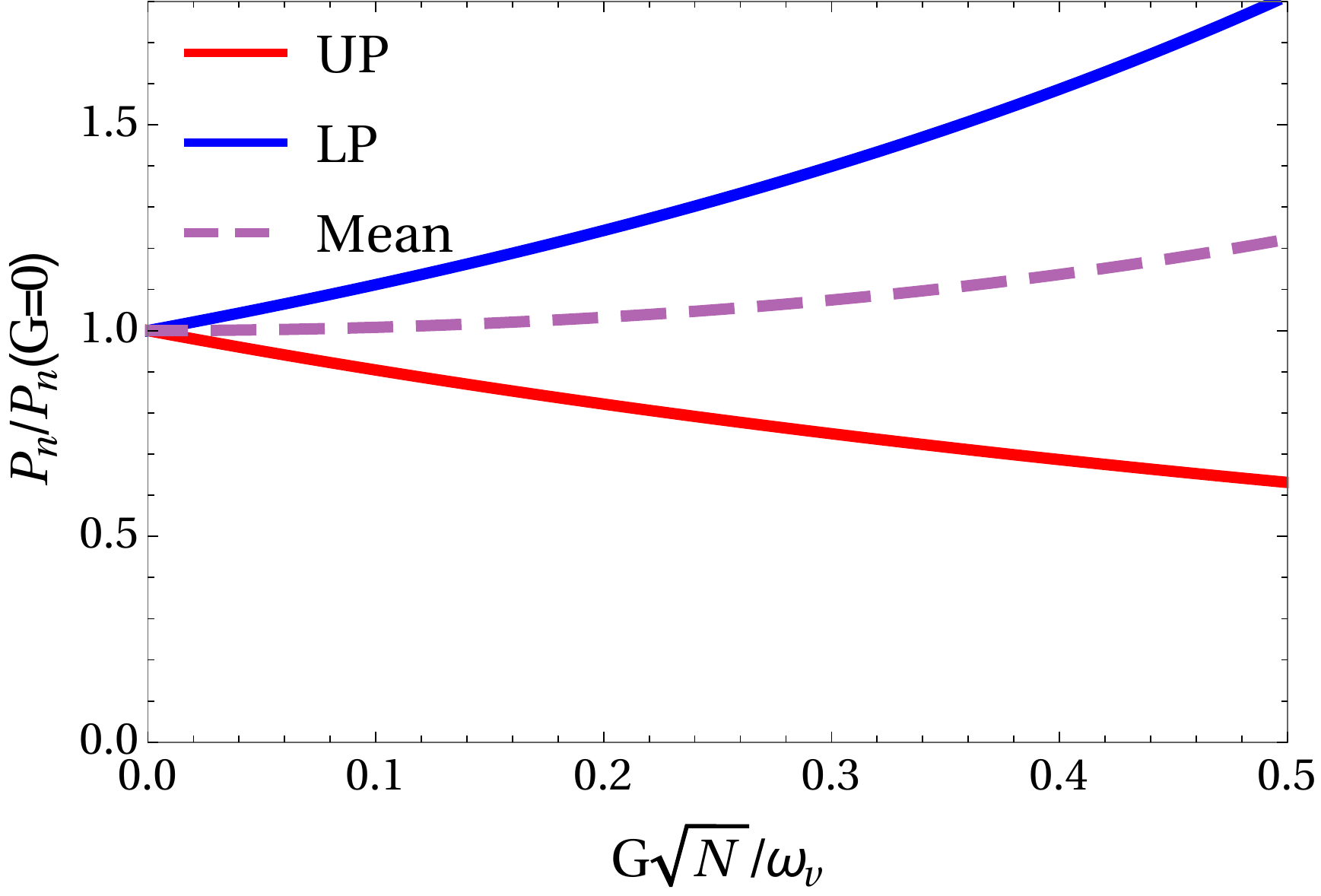}}
\caption{Transition probability to the upper and lower polariton in the RWA\@.
  Plotted for $\omega_c=\omega_v$, $S=0.3$, $N=10^6$ molecules, and $\Delta =
  \omega_v$.  Note that, as discussed in the text, such a value of $\Delta$ is
  far smaller than experimentally relevant, and thus exagerates the size of
  any effect.}
\label{RWAfig}
\end{figure}

In Figure~(\ref{RWAfig}) we plot the Raman scattering probability
(normalized by the probability in the absence of matter-light
coupling) as a function of the matter-light coupling $G$, for the
resonant case $\omega_v=\omega_c$.   For this (and subsequent) figures we choose
an unrealistically small value of $\Delta=\omega_e-\omega$, so as to
exagerate the effect of matter-light coupling, in order to see how
large the effects can be under the best possible circumstances.  We
discuss below the analytic approximation that arises for 
$\Delta \gg G\sqrt{N}, \omega_v, \omega_c$, a regime often used experimentally.
We should also note that the RWA approximation used in this section
is only valid only for $G \sqrt{N} \ll \omega_v,\omega_c$, so at the
largest values of $G\sqrt{N}$ shown, these results will be modified
as discussed below. We can however conclude that, as also found in
Ref.~\cite{pino15a}, within the limit of validity of this approach the total
Raman scattering cross-section changes only slightly with matter-light
coupling.  As one can anticipate from Eq.~(\ref{eq:ThreeModeAlpha}), on
resonance the lower polariton has a higher scattering rate due to the
larger value of $\omega_v/\omega_{LP}$.

A fully analytic result can also be extracted from this
expression by considering the limit
$\Delta \gg \omega_c, \omega_v, G\sqrt{N}$, a limit also
discussed in Ref.~\cite{pino15a}.  In this limit, the integral
over $z$ is dominated by values $z \ll 1/\Delta$, for
which one may approximate $1-e^{-z \omega_i} \simeq z \omega_i$, giving
the result:
\begin{equation}
  \label{generalapprox}
  M_k \approx  
   N
  \alpha_{k}  \omega_k
  \bigg[\Delta + \sum_{j} |\alpha_{j}|^2 \omega_j \bigg]^{-2}.
\end{equation}
For the resonant case, if we define $\zeta = G\sqrt{N}/\omega_v$ we have that
$\omega_{UP,LP} = \omega_v(1 \pm \zeta)$.  Using these expressions and
Eq.~(\ref{eq:ThreeModeAlpha}) then gives:
\begin{equation}
  \label{rwatrapprox}
  M_{k \in LP,UP} \approx   
  \frac{\sqrt{{S N}/{2}} \;\omega_v}{\left[
      \Delta + \frac{S}{N} \omega_v \left(
        N-1+ \frac{1}{1-\zeta^2}
      \right)
    \right]^2}.
\end{equation}

 Due to the $N$-dependent term in
the denominator, the effect of matter-light coupling, via $\zeta$, is
in general weak in this expression, and the upper and lower polariton
rates would be equal.    However, as $\zeta \to 1$, the
expression vanishes, as the denominator diverges.  The range of
$\zeta$ for which this divergence manifests itself is set by $1>\zeta
\gtrsim \zeta_0$, where $ \zeta_0 \simeq 1 - \frac{1}{2N}$. However,
at such strong coupling the RWA is not valid.  We will see below how
this divergence behaves beyond the RWA\@. 
In summary, for large $\Delta$, there is no enhancement of
Raman scattering within the RWA,  while for small $\Delta$,
Fig.~\ref{RWAfig} shows some enhancement.

\begin{figure}[!h]
\center{\includegraphics[width=3.2in]{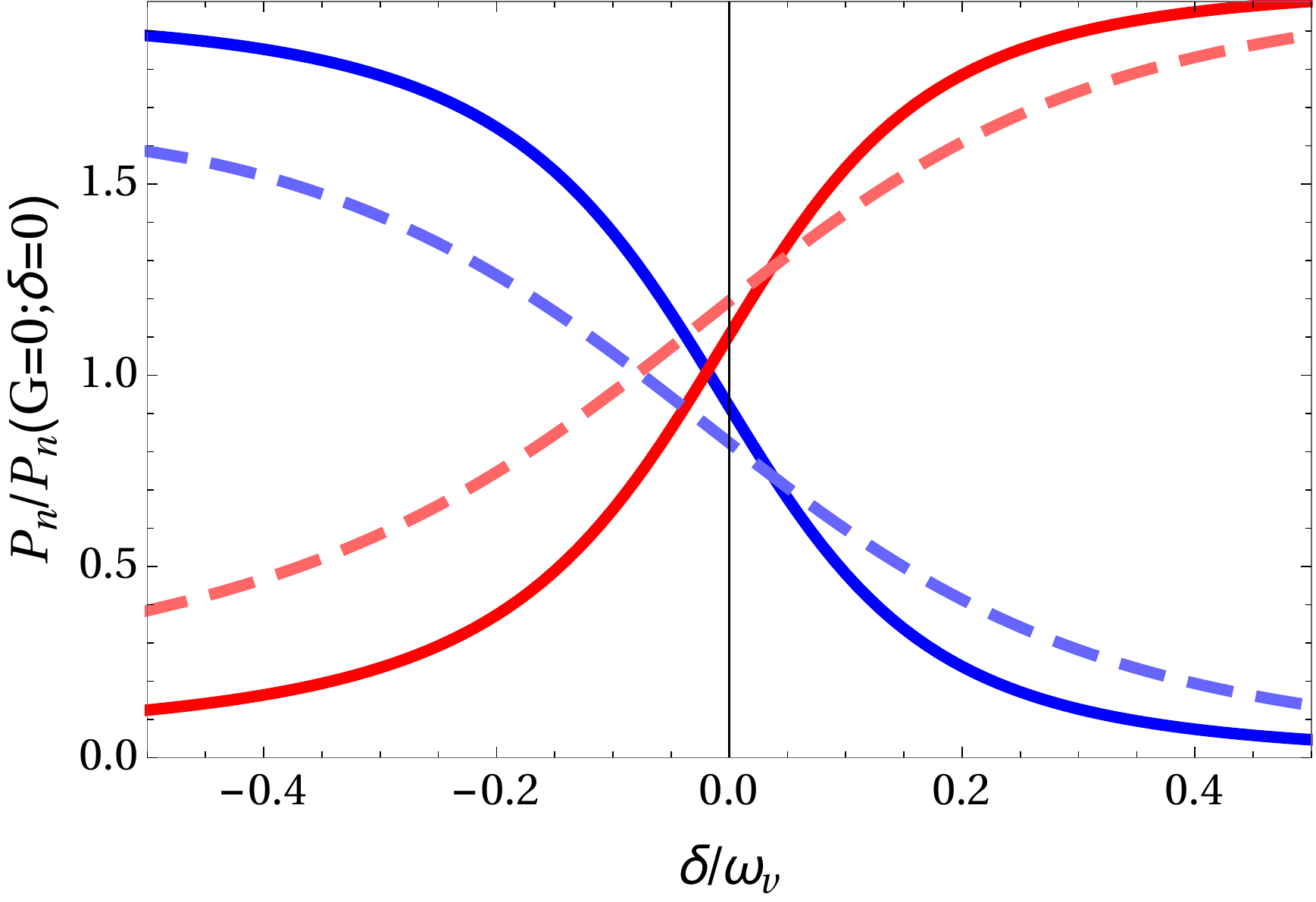}}
\caption{Transition probability to the LP and UP dependence on the
  vibron-cavity photon detuning in RWA\@. Red lines: lower polariton, blue lines: 
  upper polariton. Solid lines correspond to $G\sqrt{N} = 0.1 \omega_v$, and dashed lines to   $G\sqrt{N} = 0.2 \omega_v$.}
\label{det}
\end{figure}

In Fig.~(\ref{det}) we present the effect of the cavity--vibron
detuning $\delta \equiv (\omega_c - \omega_v)$ on the probability of
the Raman scattering. As one might expect, for large detunings the
Raman scattering occurs predominantly into the mode with the larger
excitonic component.  However, equal scattering weights require
a negative detuning, as the lower energy of the lower polariton
enhance their scattering relative to the upper polariton.

\section{Ultra-strong coupling \& $\omega_v$ dependence of the electronic state}
\label{sec:ultra-strong-coupl}

As noted earlier, in the ultra-strong coupling regime, $G\sqrt{N} \gg
\omega_v$, the RWA breaks down and we must modify the Hamiltonian in
Eq.~(\ref{hamilt1}), by replacing $G(\hat b_n^{\dagger}\hat a + \hat b_n\hat
a^{\dagger}) \to G(\hat b_n^{\dagger} + \hat b_n)(\hat a^{\dagger} + \hat a)$,
and by adding the diamagnetic $\hat A^2$ term present in the minimal coupling
Hamiltonian~\cite{cohen-tannoudji89}, i.e.\ $\frac{G^2N}{\omega_v}(\hat
a^{\dagger} + \hat a)^2$, which prevents spurious ground-state phase
transitions~\cite{rzazewski75}.  By writing the $A^2$ term in this expression
we implicitly assume the oscillator strength of the vibronic transition
is $1$, i.e.\ fully saturating the oscillator strength sum rule.  This
is a reasonable assumption for a harmonic excitation\cite{cohen-tannoudji89}.

Since the Hamiltonian no longer conserves
particle number, the intermediate and final eigenstates are no longer Fock
states.  However, as the problem remains quadratic, it can still be solved
analytically, using the position representation.  In the position
representation, we may also straightforwardly include an extra effect, missing
from Eq.~(\ref{hamilt1}), namely the possibility that the vibrational
frequency can depend on the electronic state.  The resulting Hamiltonian
including all these effects takes the form:
\begin{multline}
\label{hamilt2}
H   =  \; \omega_c \hat a^{\dagger}\hat a + 
\sum_n \biggl[ \omega_e \sigma_n^\uparrow + 
\omega_v [\hat b_n^{\dagger}\hat b^{\mathstrut}_n +  
\sqrt{S}(\hat b_n^{\dagger} + \hat b^{\mathstrut}_n)\sigma_n^\uparrow] \\ +
G(\hat b_n^{\dagger} + \hat b^{\mathstrut}_n)(\hat a^{\dagger} + \hat a) +
\nu\sigma_n^\uparrow  (\hat b_n^{\dagger} + \hat b_n^{\mathstrut})^2 
+ \frac{G^2}{\omega_c} (\hat a^{\dagger} + \hat a)^2 \biggr],
\end{multline}
where the parameter $\nu$ relates to the frequency difference $\delta
\omega_v$ between ground and excited states via $\nu = [ (\omega_v + \delta
\omega_v)^2 - \omega_v^2]/4 \omega_v$.

Before rewriting the Hamiltonian in the position representation, it is
convenient first to make a change of basis for the vibrational modes.  This
change of basis is closely related to the alternate basis for dark state modes
introduced in section~\ref{sec:alternate-basis}. However, in this case, we
make the basis change before trying to diagonalize the problem.  As seen
earlier, when molecule $m$ is excited, one can choose a basis so that $N-2$ of
the dark states do not involve any excitation of the mode $m$, and thus
decouple entirely.  In the
current context that means we choose to define $\hat b_m \to \hat b$ and $
\sum_{j \neq m} b_j/{\sqrt{N-1}} \to \hat c$.  When molecule $m$ is excited,
the remaining effective Hamiltonian can be written purely in terms of these
operators, as the other $N-2$ orthogonal modes decouple.
This then allows us to restrict our calculation of matrix elements to three
coupled harmonic oscillators.  In terms of these operators, we may write:
\begin{align}
  \label{three}
  H_{\text{eff},\Downarrow} &= \omega_c \hat a^\dagger \hat a
  + \omega_v (\hat b^\dagger \hat b + \hat c^\dagger \hat c  ) + 
  \frac{G^2 N}{\omega_c}(\hat a + \hat a^\dagger)^2
  + \nonumber\\& +
  (\hat a + \hat a^\dagger)\left(\hat b + \hat b^\dagger + \sqrt{N-1}\left( \hat c + \hat c^\dagger \right) \right)
  \\
  H_{\text{eff},m} &= H_{\text{eff},\Downarrow} + \nu (\hat b + \hat b^\dagger)^2
  + \omega_v \sqrt{S} (\hat b + \hat b^\dagger).
\end{align}
This change of basis 
does however introduce a complication when evaluating the sum over molecules,
as the labeling of final states (specifically excitations of modes $\hat b,
\hat c$) are now molecule dependent.  This can be addressed by resolving the
final state onto a fixed ``reference'' basis as is discussed further in
Appendix~\ref{sec:ThreeModes}.

\subsection{Calculating Matrix Elements}
\label{sec:calc-matr-elem-1}

To find the matrix elements between eigenstates of these Hamiltonians, we now
switch to the position representation, introducing coordinates $\hat x_i$, and
momentum $\hat p_i$ and (setting ${\hbar=1}$) such that: $\hat \psi_i =
\sqrt{\omega_i/2} (\hat x_i + i \hat p_i /\omega_i)$ for the three modes $\hat
\psi_i=(\hat a, \hat b, \hat c)$, with $\omega_i=(\omega_c, \omega_v,
\omega_v)$ respectively.  This choice of position and momentum operators means
that the problem is isotropic in momentum space, and so we can diagonalize it
by solving the classical coupled oscillator problem.  We find that both
$H_{\text{eff},\Downarrow}$ and $H_{\text{eff},m}$ can be written as
:
\begin{equation}
  \hat H_{\text{eff},\sigma} =
  \frac{1}{2} \left( \mathbf{p}^{\dagger} \mathbf{p} +
    \mathbf{x}^{\dagger} \mathbf{V}_\sigma \mathbf{x} + 2 \mathbf{h}^{\dagger}_\sigma \mathbf{x}
  \right),
\end{equation}
where we take $\sigma=\downarrow,\uparrow$ for the cases denoted as
$\Downarrow$ and $\uparrow_m$ above.  The matrices and vectors
appearing here are then $\mathbf{h}_\downarrow=0, \mathbf{h}_\uparrow = \left( 0, \omega_v \sqrt{2 \omega_v S}, 0  \right)^T$ and 
\begin{align*}
  \mathbf{V}_{\downarrow} &=
  \begin{pmatrix}
    \omega_c^2 + 4 G^2 N & \xi & \xi\sqrt{N-1} \\
    \xi & \omega_v^2 & 0 \\
    \xi\sqrt{N-1} & 0 & \omega_v^2
  \end{pmatrix}
  \\
  \mathbf{V}_{\uparrow} &= \mathbf{V}_{\downarrow} +
  \begin{pmatrix}
    0 & 0 & 0 \\
    0 & 4\varepsilon_v \nu & 0 \\
    0 & 0 & 0
  \end{pmatrix},
\end{align*}
and we introduced the shorthand $\xi = 2G\sqrt{\omega_v\omega_c}$.

We can clearly diagonalize $\hat H_\sigma$ by writing: $\mathbf{x} =
\mathbf{U}_\sigma \mathbf{X}_\sigma - \mathbf{V}_\sigma^{-1} \mathbf{h}_\sigma
$ where $\mathbf{U}^\dagger_\sigma \mathbf{V}_\sigma \mathbf{U}_\sigma \equiv
\mathbf{\Omega}^2_{\sigma}$ is diagonal. Note that $\mathbf{V}$ is a real
symmetric matrix, and so although we write Hermitian conjugates, these are all
equivalent to transposes. After diagonalization
one finds $H_{\text{eff},\sigma} =\frac{1}{2} \sum_i (
{P}^{2}_{i,\sigma} + {\Omega}_{i,\sigma}^2 {X}^{2}_{i,\sigma} ) + \text{const.}$,  
thus, one can write eigenfunctions
in the position basis as:
\begin{multline}
\Psi_{l_1l_2l_3,\sigma}(x_a,x_b,x_c) =
   \sqrt[4]{\Omega_{i,\sigma}\Omega_{2,\sigma}\Omega_{3,\sigma}}
  \psi_{l_1}\left(X_{1,\sigma} \sqrt{\Omega_{1,\sigma}} \right) \\
  \psi_{l_2}\left(X_{2,\sigma} \sqrt{\Omega_{2,\sigma}} \right)
  \psi_{l_3}\left(X_{3,\sigma} \sqrt{\Omega_{3,\sigma}} \right),
\end{multline}
where $\psi_l(y)$ are the Gauss-Hermite functions
\begin{displaymath}
  \psi_l(y) =
  \frac{1}{\sqrt{\sqrt{\pi} 2^l l!}} H_l(y) e^{-y^2/2},
\end{displaymath}
$\Omega_{i\sigma}$ are the diagonal elements of $\mathbf{\Omega}_\sigma$, and
the components $X_i$ are related to $x_i$ by the linear transformation given
above.

Now, as in Eq.~(\ref{mk}), we need to calculate $M_k$, which involves a sum of
transition matrix elements over all intermediate states, divided by
corresponding energy differences.  The transition matrix elements can be
written using position basis overlaps of eigenfunctions.  Using the wavefunctions introduced above and rewriting the 
denominator as an integral over $z$ as before, we get that the
matrix element to a final state with mode $k$ excited is:
\begin{multline}
\label{summ}
M_k  = N
\sqrt{2 \Omega_{k,\downarrow} 
  \prod_i \left(\Omega_{i,\uparrow}\Omega_{i,\downarrow}\right)}
\int_0^{\infty} ds e^{-s \Delta}
\int d^3x d^3x' 
\\
\prod_i \Biggl[
\sum_{l_i}  
\psi_{l_i} \left( \sqrt{\Omega_{i,\uparrow} } X_{i,\uparrow} \right)
\psi_{l_i} \left( \sqrt{\Omega_{i,\uparrow} } X^\prime_{i,\uparrow} \right)
e^{-s l_i \Omega_{i,\uparrow}} 
\\ 
\times
\psi_0 \left( \sqrt{\Omega_{i,\downarrow} } X_{i,\downarrow} \right)
\psi_0 \left( \sqrt{\Omega_{i,\downarrow} } X^\prime_{i,\downarrow} \right) 
\Biggr]
X_{k,\downarrow}.
\end{multline}
In writing the above, we have used the fact that for bright modes, the sum
over molecules is replaced by a factor $N$, while for dark modes the sum over
molecules vanishes (see Appendix~\ref{sec:ThreeModes}).  We have also used the
fact that the first-excited Hermite mode is related to the ground state by
$\psi_1(y) = \psi_0(y) \sqrt{2} y$.

To calculate the coordinate integrals in Eq.~(\ref{summ}) we may first note
that since $ \mathbf{x}, \mathbf{X}_\sigma$ are all related by unitary
transformations, we can change the integration coordinates to $\mathbf{X}_{i\uparrow}$
with unit Jacobian.   The resulting integral then involves known
overlaps of Gauss-Hermite functions.  For further details, see
Appendix~\ref{HermiteIntegrals}. The result is
\begin{multline}
  \label{final}
  M_k  = 8N \sqrt{2 \Omega_{k,\downarrow}} 
  \left[ \mathbf{U}_{\downarrow}^{\dagger}\mathbf{U}_{\uparrow} \right]_{kr}
  \\ \times
  \int ds e^{-s \Delta}  
  \prod_i \left( \sqrt{\frac{\Omega_{i,\downarrow}\Omega_{i,\uparrow}}{1 - \exp(-2s \Omega_{i,\uparrow})}} \right)  
  \\ \times
  \frac{(\mathbf{A}^{-1}\mathbf{q} - \mathbf{l})_r}{\sqrt{\det(\mathbf{A})}}
  \exp\left[ \frac 1 2 \mathbf{q}^\intercal \mathbf{A}^{-1} \mathbf{q} - \mathbf{l}^\intercal \mathbf{R} \mathbf{l} \right],
\end{multline}
where we have introduced the $6\times 6$ matrix $\mathbf{A}$ which naturally
comes after computing the six dimensional Gaussian integrals in Eq.~(\ref{summ}).  This matrix
can be written in block form as:
\begin{equation}
  \label{Amatrix}
  \mathbf{A} =
  \begin{pmatrix} 
    \mathbf{P} + \mathbf{R} & - \mathbf{Q} \\
     - \mathbf{Q} & \mathbf{P} + \mathbf{R} 
  \end{pmatrix}
\end{equation}
where the $3 \times 3$ blocks are given by $\mathbf{R} = \mathbf{U}_{\uparrow}^\intercal
\mathbf{U}_{\downarrow} \mathbf{\Omega}_{\downarrow}
\mathbf{U}_{\downarrow}^\intercal \mathbf{U}_{\uparrow}$, $\mathbf{P} =
\diag\left(\frac{\Omega_{i,\uparrow}}{\tanh(s\Omega_{i,\uparrow})} \right)$,
and ${\mathbf{Q} =
  \diag\left(\frac{\Omega_{i,\uparrow}}{\sinh(s\Omega_{i,\uparrow})}\right)}$.
The three- and six-component vectors appearing in Eq.~(\ref{final}) are given by
$\mathbf{l} =
\mathbf{\Omega}_{\uparrow}^{-2}\mathbf{U}_{\uparrow}^{\dagger}\mathbf{h}_{\uparrow}$,
and $\mathbf{q}^\intercal = (\mathbf{l}^\intercal \mathbf{R}, \mathbf{l}^\intercal \mathbf{R})$.  This is as far as we can simplify this expression in
the general case, where $\nu \neq 0$, but Eq.~(\ref{final}) can nonetheless
be evaluated efficiently numerically.

\subsection{Numerical results and large $\Delta$ approximation}
\label{sec:numer-results-large}

\begin{figure}[!h]
\center{\includegraphics[width=3.2in]{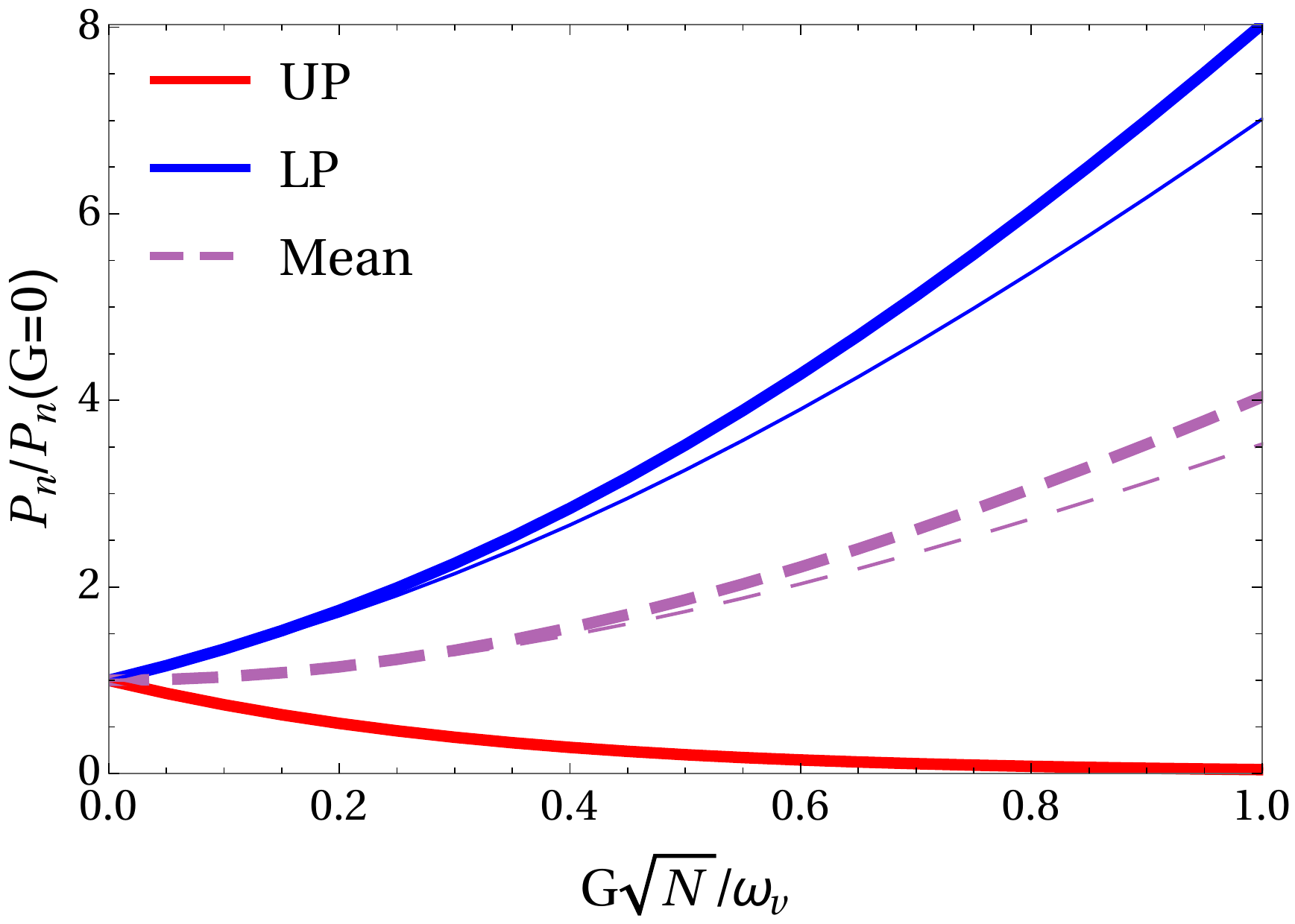}}
\caption{Transition probability to the upper and lower polariton in beyond the
  RWA, including $A^2$ terms.  Thick lines plotted for $\delta\omega_v=0$,
  thin lines to $\delta\omega_v=-0.5 \, \omega_v$. Other parameters as for
  Fig.~\ref{RWAfig}. }
\label{non_RWA_Asq}
\end{figure}

In Figure~\ref{non_RWA_Asq} we compare the behavior with and without
electronic-state dependent vibrational frequency.  It is clear the inclusion
of this term makes only minor changes. It is worth noting that while
the detuning $\nu$ mixes bright and dark states in the excited state manifold,
there is no such mixing in the final (electronic ground state manifold).
Thus, the effect of $\nu$ is only to modify the intermediate states
appearing in the calculation of the transition amplitude.

On the other hand, as we will discuss next, the correct treatment of the
ultra-strong coupling (including the diamagnetic terms) has a significant
effect, avoiding features associated with the ground state phase transition.

Since the electronic state dependence of vibrational frequency is unimportant,
we may focus on the case $\nu=0$.  In this case, Eq.~(\ref{final}) simplifies
considerably, as we have $\mathbf{V}_{\uparrow} = \mathbf{V}_{\downarrow}$,
and so consequently $\mathbf{U}_{\uparrow} = \mathbf{U}_{\downarrow}$ and
$\mathbf{\Omega}_{\uparrow} = \mathbf{\Omega}_{\downarrow}$.  This then
in turn means that $\mathbf{R}=\mathbf{\Omega}$ becomes diagonal, and
so the matrix $\mathbf{A}$ can be rewritten as three $2\times 2$ blocks,
and thus inverted in closed form.
After some algebra, this leads to an
expression of exactly the same form as (\ref{MkRWA}), but with the three
coefficients $\alpha_i$ given by $\alpha_i = l_i \sqrt{\Omega_i/2}$.
In the resonant case $\omega_c=\omega_v$ this simplifies further to:
\begin{equation}
  \label{alphaNonRWA}
  \alpha_i = \sqrt{\frac{S}{N}}  \left(
    \frac{1}{\sqrt{2}}\frac{\omega_v^{3/2}}{\omega_{LP}^{3/2}},
    \frac{1}{\sqrt{2}}\frac{\omega_v^{3/2}}{\omega_{LP}^{3/2}},
    {\sqrt{N-1}},
\right).
\end{equation}

As discussed in Section~\ref{sec:numerical-results}, the asymptotic behavior
at large $\Delta$ has a simple form.  Using Eq.~(\ref{generalapprox}) we
now have that the large $\Delta$ asymptote of the resonant
case gives
\begin{equation}
\label{singexpr}
  M_{k=LP,UP}
  \approx
  \frac{%
  \sqrt{\frac{SN}{2} 
    \frac{\omega_v^3}{\omega_{k}}}}
  {\left[ 
      \Delta + 
      \frac{S}{N} \omega_v
      \left(
        N-1 +
        \frac{1}{2}\left[
        \frac{\omega^2_v}{\omega_{UP}^{2}} 
        + 
        \frac{\omega_v^2}{\omega_{LP}^{2}}  
      \right]
    \right)  
  \right]^2},
\end{equation}
Note that in contrast to Eq.~(\ref{rwatrapprox}), the numerator retains a
dependence on $\omega_k$, due to the extra powers of $\omega_k$ in the
definition of $\alpha_k$.  Thus, as the lower polariton frequency tends to
zero with increasing coupling, the numerator will diverge.  This
means that beyond the RWA, even for large $\Delta$, there is a
growth of Raman scattering with $G$.  This was also seen by \citet{pino15a}
for a single molecule. However, at very
strong coupling one once again has a divergence of the denominator that is
stronger than that of the numerator.  Thus the asymptotic limit of strong
coupling is in fact for the expression to vanish. This can be seen
most clearly by again using $\zeta = G\sqrt{N}/\omega_v$.  Writing
the eigenfrequencies
$\omega_{UP,LP}^2 = \omega_v^2(1 + 2 \zeta^2 \pm 2 \zeta \sqrt{1+\zeta^2})$
this yields:
\begin{equation}
\label{singexprzeta}
  M_{k=LP,UP}
  \approx
  \frac{\sqrt{{SN}/{2}} \; \omega_v  \sqrt{\omega_v/\omega_k}}
  {\left[ 
      \Delta + 
      \frac{S}{N} \omega_v
      \left(
        N+ 2 \zeta^2
    \right)  
  \right]^2},
\end{equation}
At large $\zeta$ one has $\omega_{LP} \simeq \omega_v/2\zeta$, making
the relatively scaling of numerator and denominator clear.   Note however
that for $\zeta^2$ to dominate the denominator would require
the (currently unattainable) limit $G \gg \omega_v$, i.e.\ ultra-strong
single-molecule coupling.

This expression also shows the crucial role played by the $A^2$ term at
ultra-strong coupling.  Unlike the rotating wave approximation, where
$\omega_{LP}$ diverges as $\zeta \to 1$, here the LP energy always remains 
finite (there is no superradiance transition~\cite{rzazewski75}), and instead
leading to the LP energy vanishing asymptotically at $\zeta \to \infty$.    As
such, the Raman scattering probability is a smooth function of the coupling
strength and neither vanishes nor diverges at any finite coupling strength.
One should however note that the assumption
$\omega_{UP} \ll \Delta$ required to make the large
$\Delta$ expansion in Eq.~(\ref{singexpr},\ref{singexprzeta}) will fail
in the limit $\zeta \to \infty$.  In this limit one must therefore
return to using Eq.~(\ref{MkRWA},\ref{alphaNonRWA}).

\section{Multiple excitations}
\label{sec:multiple-excitations}

So far we have determined the Raman transition amplitudes to final states with
a single upper or lower polariton.  In this section, we discuss how the
tractable expressions we derived above for transition matrix elements can also
be extended to multiple excitations.  Specifically, we consider the RWA
expression for the transition amplitude to a state where mode $i$ has $q_i$
excitations.  Details of the calculation are given in
Appendix~\ref{sec:multiple-final-state}.  The compact expression for this is
given by: 
\begin{multline}
\label{multpoltransit}
  M_{\{q_i\}} = 
  \sum_{m} \int\limits_0^{\infty}dz e^{-z \Delta} 
  \prod_i
  (\alpha_{m,i}^{\ast})^{q_i}
  \frac{(e^{-z \omega_i}-1)^{q_i}}{\sqrt{q_i!}}
  \\
  \times
  \exp\bigl[ - |\alpha_{m,i}|^2(1-e^{-z\omega_i})\bigr].
\end{multline}
One can immediately see that if $q_{i=k}=1, q_{i\neq k}=0$, this reduces
to the formula given in Eq.~(\ref{MkRWA}).

If we consider the special case where a single mode is multiply occupied, so
$q_{i=LP}=n$, $q_{i \neq LP}=0$, the
formula simplifies as all terms are molecule independent and so $\sum_m \to
N$.  In this case we can see that the transition amplitude to the multiple
lower polariton state has a stronger dependence on $\omega_{LP}$, increasing
as $\omega_{LP}^{-n}$, as might be expected from multiplying together the
amplitudes for $n$ excitations.  However, the scaling with number of molecules
is different: The expression for transition amplitude to $n$ lower polaritons
is proportional to $N^{1-n/2}$.  i.e., while the Raman transition \emph{probability} to one-excitation final states scales as $N$, the transition probability
to two-excitation final states does not scale with $N$.
It is however important to note that within the
multiple excitation sector, other final states are possible.  For example, a
Raman transition to dark modes can now occur: if one considers modes $k,
k^\prime$ using the basis choice of Eq.~(\ref{UmatStd}), such that
$k+k^\prime=N$, then one may see that $\sum_m \alpha_{k,m} \alpha_{k^\prime,m}
\neq 0$.  i.e., ``momentum'' conserving pairs of dark modes become possible.
As such, the total transition probability to all two-excitation final states
scales as $N$, the same scaling as single-excitation final states.
However, the two-excitation final states are dominated by the dark
state pairs.

\section{Conclusion}
\label{sec:conclusion}

In this work we analyzed the effect of strong photon--vibron coupling on the
Raman scattering intensity, and show that a compact analytic expression can be
found for the Raman transition amplitude.  As also found in Ref.~\cite{pino15a}, we
find that matter-light coupling leads to a redistribution of the vibronic
Raman signal between upper and lower polariton modes (and no scattering into
single dark states).  At leading order in matter-light coupling, there is no
change to the overall scattering amplitude, but changes do occur at higher
orders.  Under ultra-strong coupling we see significant enhancement of the
scattering into the lower polariton due to the mode softening, and suppression
of Raman scattering into the upper polariton, so the overall signal goes up.
In considering this ultra strong coupling limit, we showed that $A^2$ terms
are essential in preventing (unphysical) divergence of the Raman signal at
finite coupling strength.  In contrast, we find that
electronic-state-dependent vibronic frequency shifts have a negligible effect
of the Raman scattering amplitude.  We also showed that for Raman scattering
to sectors with multiple excitations show a system-size suppression of
individual matrix elements, however transitions to states involving multiple
dark states now become possible.

\begin{acknowledgments}
  AS acknowledges support from the EPSRC CM-CDT (EP/L015110/1). JK
  acknowledges financial support from EPSRC program ``Hybrid
  Polaritonics'' (EP/M025330/1) and from the Leverhulme Trust
  (IAF-2014-025).  We are grateful to J. Feist and colleagues
  for helpful comments on an earlier version of this manuscript.
\end{acknowledgments}

\appendix

\section{Dark modes and three mode description}
\label{sec:ThreeModes}

This appendix addresses a subtle issue about considering transition matrix
elements in the ``three mode'' basis $\hat a, \hat b_m \to \hat b, \sum_{j
  \neq m} b_j/{\sqrt{N-1}} \to \hat c$ used in writing Eq.~(\ref{three}).  The
issue is that the set of eigenmodes then used to describe the dark states is
dependent on which molecule is excited.  Since the overall transition matrix
element requires summing over molecules, some care is required to correctly
perform this sum and see that dark states still cancel.  In contrast the
bright states pose no issues, since the bright states are non degenerate, and
so uniquely determined independent of basis --- the issue with dark states is
that degeneracy allows us freedom to choose the set of states, and our three
mode basis chooses a different set of eigenmodes for each molecule.

For the single excitation final state that we consider throughout
most of the paper, it is clearest to use a first-quantized Dirac
notation to discuss the issue.  Our three modes can be considered as the
cavity mode $|\psi_a \rangle = |1;0,0,\ldots0\rangle$, and the two vibronic
modes $|\psi^{(m)}_b \rangle = |0;\ldots,0,1,0,\ldots\rangle$, and
$|\psi_c^{(m)} \rangle = |0;\ldots,1,0,1,\ldots\rangle/\sqrt{N-1}$ where the
non-zero (zero) element in mode b (c) corresponds to the excited
molecule $m$.  In terms of these basis states, the eigenmodes are the two
polaritonic states and the dark states are:
\begin{align*}
| LP, UP \rangle &= |\psi_a \rangle \pm \frac{1}{\sqrt{N}} \left( |\psi_b^{(m)} \rangle + \sqrt{N-1}|\psi_c^{(m)} \rangle \right),
\\
| D^{(m)} \rangle &=\frac{1}{\sqrt{N}} \left( \sqrt{N-1} |\psi_b^{(m)} \rangle - |\psi_c^{(m)} \rangle \right).
\end{align*}

In order to correctly sum the contributions of transition amplitudes to the
states $|D^{(m)}\rangle$ for different molecules, we should resolve these
states onto a fixed reference state.  i.e., we should define a (dark) state
$|X\rangle$ and calculate the transition probability $P_X \propto | \sum_m
\langle X| D^{(m)} \rangle M^{(m)}_{k=D^{(m)}}|^2$ where $M^{(m)}_k$ is the
transition amplitude coming from excitations of molecule $m$.  This is the
correct way to deal with sum over molecules appearing in the Raman transition
amplitude.

With this expression, we can indeed show that the total dark state probability
vanishes.  Suppose we take as our reference $| X \rangle = | D^{(1)} \rangle$.
The overlaps required then involve the need to use the overlap:
\begin{displaymath}
  \langle D^{(1)} | D^{(m \neq 1)} \rangle = 
  \frac{1}{N} \left( -2  + \frac{N-2}{N-1} \right)
  =
  - \frac{1}{N-1}.
\end{displaymath}
From our calculation in section~\ref{sec:ultra-strong-coupl}, we find that
$M^{(m)}_{k=D^{(m)}}$ is independent of molecule label $m$, so we find that
$P_X \propto |\sum_m \langle X | D^{(m)} \rangle|^2 = 0$. This demonstrates
again that the amplitude for transition to dark modes vanishes, and confirms
that we may use such a basis to simplify the calculations, as used in
Sec.~\ref{sec:ultra-strong-coupl}.

\section{Details of non RWA calculation}
\label{HermiteIntegrals}

This appendix provides further details of the steps required to evaluate the
sums over modes and 
Gaussian integrals in Eq.~(\ref{summ}).  As noted in
section~\ref{sec:ultra-strong-coupl}, since the
Jacobian for an unitary transformation is $1$, we may choose to write the
integrals in terms of the variables $\mathbf{X}_{\uparrow},
\mathbf{X}_{\uparrow}^\prime$.  It is convenient to denote $\mathbf{X} \equiv
\mathbf{X}_{\uparrow}$ in terms of which
\begin{equation}
\mathbf{X}_{\downarrow} = \mathbf{U}_{\downarrow}^{\dagger} \left( \mathbf{U}_{\uparrow}\mathbf{X} - \mathbf{V}_{\uparrow}^{-1} \mathbf{h}_{\uparrow} \right) = 
\mathbf{U}_{\downarrow}^{\dagger} \mathbf{U}_{\uparrow} \left( \mathbf{X} - \mathbf{l} \right),
\end{equation}
where we introduced
$
\mathbf{l} = \mathbf{U}_{\uparrow}^{\dagger}\mathbf{V}_{\uparrow}^{-1}\mathbf{h}_{\uparrow} = \mathbf{\Omega}_{\uparrow}^{-2}\mathbf{U}_{\uparrow}^{\dagger}\mathbf{h}_{\uparrow}.
$
A similar set of relations hold for the primed coordinates.  

\begin{widetext}
The sum over modes can be evaluated using a version of Mehler's formula
\footnote{This formulae is frequently known in the context
of the imaginary time Green's function for an harmonic oscillator}:
\begin{displaymath}
  \sum_l \psi_l \left( \sqrt{\Omega } X \right)
  \psi_l \left( \sqrt{\Omega } X^\prime \right)  e^{-s l \Omega} 
   =
  \frac{1}{\sqrt{\pi}} \frac{1}{\sqrt{1 - e^{-2 s \Omega}}}
  \exp\left[ - \frac{\Omega}{2} \left(
      \frac{X^2 + X^{\prime\;2}}{\tanh(s \Omega)} - \frac{2 X X^\prime}{\sinh(s \Omega)}
    \right)
  \right].
\end{displaymath}
 With this expression Eq.~(\ref{summ}) can be reduced to:
\begin{multline}
  \label{nonRWAMKIntegral}
  M_k = \frac{\sqrt{2 \Omega_{k, \downarrow}} }{\pi^3}
  \left[ \mathbf{U}_{\downarrow}^{\dagger}\mathbf{U}_{\uparrow} \right]_{kr}
  \int ds e^{-s \Delta} 
  \int \prod_i \left( d^3 X_i d^3 X^\prime_i 
    \frac{\sqrt{\Omega_{i,\uparrow} \Omega_{i,\downarrow}}}{%
      \sqrt{1 - e^{-2 s \Omega_{i,\uparrow}}}} \right)  
    ({X}_r - {l}_r)
\\  \exp
  \left[
  - \frac 1 2
  \left(
  (\mathbf{X}-\mathbf{l})^\intercal 
  %\mathbf{U}_{\uparrow}^\intercal \mathbf{U}^{\mathstrut}_{\downarrow} 
  %\mathbf{\Omega}^{\mathstrut}_{\downarrow} 
  %\mathbf{U}_{\downarrow}^\intercal \mathbf{U}^{\mathstrut}_{\uparrow} 
  \mathbf{R}
  (\mathbf{X} - \mathbf{l}) + 
  (\mathbf{X}^\prime-\mathbf{l})^\intercal 
  \mathbf{R}
%  \mathbf{U}_{\uparrow}^\intercal \mathbf{U}^{\mathstrut}_{\downarrow} 
%  \mathbf{\Omega}^{\mathstrut}_{\downarrow} 
%  \mathbf{U}_{\downarrow}^\intercal \mathbf{U}^{\mathstrut}_{\uparrow} 
  (\mathbf{X}^\prime - \mathbf{l})
  \right)
 - \sum_i \frac{\Omega_{i,\uparrow}}{2} \left(
      \frac{X_i^2 + X_i^{\prime\;2}}{\tanh(s \, \Omega_{i,\uparrow})} - \frac{2 X_i X_i^\prime}{\sinh(s \, \Omega_{i,\uparrow})}
    \right)
  \right] 
\end{multline}
where we introduced the matrix
$\mathbf{R} \equiv \mathbf{U}_{\uparrow}^\intercal \mathbf{U}^{\mathstrut}_{\downarrow} \mathbf{\Omega}^{\mathstrut}_{\downarrow} \mathbf{U}_{\downarrow}^\intercal \mathbf{U}^{\mathstrut}_{\uparrow}$.
\end{widetext}

Equation~(\ref{nonRWAMKIntegral}) involves a Gaussian integral over the six components $X_i,
X_i^\prime$ which we may define as $O_r(s)$, such that
\begin{multline*}
M_k=
  \frac{\sqrt{2 \Omega_{k, \downarrow}} }{\pi^3}
  \left[ \mathbf{U}_{\downarrow}^{\dagger}\mathbf{U}_{\uparrow} \right]_{kr}
  \\\times\int ds e^{-s \Delta} \prod_i
  \left(
    \sqrt{\frac{\Omega_{i,\uparrow}\Omega_{i,\downarrow}}{1 - e^{-2s \Omega_{i,\uparrow}}}}
  \right) O_r(s),
\end{multline*}
To proceed further, we can notice that $O_r(s)$ is a 6 dimensional Gaussian
integral, and so can be calculated analytically. Defining the six dimensional
coordinates: $\mathbf{z}^\intercal \equiv
(\mathbf{X}^\intercal,\mathbf{X}^{\prime \intercal})$ the
Gaussian integral $O_r(s)$ can be written as:
\begin{displaymath}
O_r(s) = \int d^6 z (z_r - l_r) \exp\left[ -\frac 1 2 \mathbf{z}^\intercal \mathbf{A} \mathbf{z} + \mathbf{q}^\intercal \mathbf{z} - c \right],
\end{displaymath}
where the matrix $6\times 6$ matrix $\mathbf{A}$ is as given in Eq.~(\ref{Amatrix}),
the $6$ component vector $\mathbf{q}^\intercal = (\mathbf{l}^\intercal
\mathbf{R}, \mathbf{l}^\intercal \mathbf{R})$, and the constant $c =
\mathbf{l}^\intercal \mathbf{R} \mathbf{l}$. Thus, computing the Gaussian
integral over coordinates $\mathbf{z}$, we eventually obtain
\begin{displaymath}
\mathbf{O}_r(s) =  \frac{(2\pi)^3 \left( \mathbf{A}^{-1}\mathbf{q} - \mathbf{l} \right)_r}{\sqrt{\det(\mathbf{A})}} \exp\left[ \frac 1 2 \mathbf{q}^\intercal \mathbf{A}^{-1}\mathbf{q} - \mathbf{l}^\intercal \mathbf{R} \mathbf{l} \right],
\end{displaymath}
and so derive the final expression Eq.~(\ref{final}).

\section{Multiple final-state  excitations}
\label{sec:multiple-final-state}
This appendix provides further details on how to calculate the transition rate
to a final state with multiple excitations.  For simplicity, we present the
result as can be calculated in the rotating wave approximation.  In this case,
we can use the Fock state basis, as discussed in
Sec.~\ref{sec:calc-matr-elem}.  If we consider the state in which mode $i$ has
$q_i$ excitations, we must replace the matrix element between intermediate and
final states in Eq.~(\ref{oneExME}) with one describing transitions to a final
state specified by the occupations $\{q_i\}$.  By considering the
combinatoric factors associated with the overlap between $\{p_i \}$
(displaced) excitations in the intermediate state and $\{q_i\}$ in the final
state, one may show that:
\begin{multline*}
\mathpzc{M}_{\{q_i\},\{p_i\}}^{(m)}
=
\mathpzc{M}_{0,\{p_i\}}^{(m)}
\\
\times\prod_i \sqrt{q_i!}
\sum_{l_i=q_i-p_i}^{q_i}\!\!\!
\frac{(-1)^{l_i}}{l_i!}
\frac{|\alpha_{m,i}|^{2l_i}}{\alpha_{m,i}^{\ast\; q_i}}
{{p_i}\choose{q_i-l_i}}
\end{multline*}
where the last term is the binomial coefficient.
With this result, the Raman transition amplitude $M_{\{q_k\}}$ can be
written using the same exponentiation of denominator as used previously, to
give:
\begin{multline*}
  M_{\{q_i\}} = 
  \int\limits_0^{\infty}dz e^{-z \Delta} \sum_{m} 
  \prod_i \sqrt{q_i!} \frac{e^{-|\alpha_{m,i}|^2}}{\alpha_{m,k}^{q_i}}
  \\\times
  \sum_{l_i=0}^{q_i} \sum_{p_i=q_i-l_i}^{\infty}
  \frac{(-1)^{l_i}   |\alpha_{m,i}|^{2(l_i+ p_i)}
  e^{- p_i z \Omega_i}}{l_i! (p_i-q_i+l_i)!(q_i-l_i)!},
\end{multline*}
where we have swapped the order of summation over $l_i$ and $p_i$.
One may then identify the sum over $p_i$ as being the Taylor
expansion of an exponential, to give
\begin{multline*}
  M_{\{q_i\}} = 
  \int\limits_0^{\infty}dz e^{-z \Delta} \sum_{m} 
  \prod_i \sqrt{q_i!} 
  \frac{e^{-|\alpha_{m,i}|^2}(1-e^{-z\Omega_i})}{\alpha_{m,k}^{q_i}}
  \\\times
  |\alpha_{m,i}|^{2 q_i}  \sum_{l_i=0}^{q_i} 
  \frac{(-1)^{l_i}   
    e^{-z \Omega_i (q_i-l_i)}}{l_i! (q_i-l_i)!},
\end{multline*}
and then performing the sum over $l_i$ gives the expression 
in Eq.~(\ref{multpoltransit}).

\bibliography{literature}

\end{document}